\input harvmac

\def\Title#1#2{\rightline{#1}\ifx\answ\bigans\nopagenumbers\pageno0
\vskip0.5in
\else\pageno1\vskip.5in\fi \centerline{\titlefont #2}\vskip .3in}

\font\caps=cmcsc10

\noblackbox
\parskip=1.5mm

  
\def\npb#1#2#3{{\it Nucl. Phys.} {\bf B#1} (#2) #3 }

\def\bb#1{{\tt hep-th/#1}}


\def\CL{{\cal L}}


\def\dj{\hbox{d\kern-0.347em \vrule width 0.3em height 1.252ex depth
-1.21ex \kern 0.051em}}

\def\half{{1\over 2}\,}

\def\ket{\rangle}

\def\epb{\overline \varepsilon}   
\def\ep{\varepsilon}    
\lref\regk{S. Elitzur, A. Giveon and D. Kutasov,    {\it``Branes and
$N=1$ Duality in String Theory"}, Weizmann preprint 
WIS/97/6, RI-1-97, \bb{9702014.}}   
\lref\rangles{M. Berkooz, M.R. Douglas and R.G. Leigh, \npb {480}{1996}
{265,} \bb{9606139\semi}
V. Balasubramanian and R.G. Leigh, {\it ``D-Branes, Moduli and Supersymmetry",
} Princeton Univ. preprint PUPT-1654, \bb{9611165.}}   
\lref\rsei{N. Seiberg, \npb {435}{1995}{129,} \bb{9411149.}} 
\lref\rvaf{M. Bershadsky, A. Johansen, T. Pantev, V. Sadov and
C. Vafa, {\it ``F-theory, Geometric Engineering and $N=1$ dualities",}
 Harvard preprint HUTP-96/A057, 
 \bb{9612052\semi} C. Vafa and B. Zwiebach, {\it ``$N=1$ Dualities of
$SO$  and $Usp$ Gauge Theories and T-Duality of String Theory",}
Harvard preprint HUTP-97/A001,  \bb{9701015\semi}
K. Hori and Y. Oz, {\it ``F-Theory, T-Duality on $K3$ Surfaces and
$N=2$ Supersymmetric Gauge Theories in Four Dimensions", }
Berkeley preprint  LBNL-40031, \bb{9702173.} }
\lref\rvafo{C. Vafa and H. Ooguri, {\it ``Geometry of $N=1$ dualities
in Four Dimensions",} Harvard preprint HUPT-97/A010,  \bb{9702180.}}
\lref\rhw{A. Hanany and E. Witten, {\it ``Type-IIB Superstrings, BPS
Monopoles, And Three Dimensional Gauge Dynamics",} IAS, Princeton
  preprint
IASSNS-HEP-96/121,  \bb{9611230.}} 
\lref\rdosdual{I. Antoniadis and B. Pioline, {\it ``Higgs branch, 
Hyper-Kahler quotient and duality in SUSY $N=2$ Yang--Mills Theories",}
Ecole Polytechnique preprint CPTH-S459-0796,  \bb{9607058.}}  
\lref\rdeff{P.C. Argyres, M.R. Plesser and N. Seiberg, \npb{471}{1996}
{159,}  \bb{9603042.}}


\line{\hfill CERN-TH/97-38}
\line{\hfill {\tt hep-th/9703051}}
\vskip 0.5cm

\Title{\vbox{\baselineskip 12pt\hbox{}
 }}
{\vbox {\centerline{Rotated Branes and $N=1$ Duality  }
}}

\centerline{$\quad$ {\caps J. L. F. Barb\'on
 }}
\smallskip

\centerline{{\sl Theory Division, CERN}}
\centerline{{\sl 1211 Geneva 23, Switzerland}}
\centerline{{\tt barbon@mail.cern.ch}}

 \vskip 1.0in

We consider configurations of rotated $NS$-branes leading     
to a family of four-dimensional $N=1$ super-QCD 
 theories, interpolating  
between  four-dimensional analogues of the Hanany-Witten vacua, 
 and the Elitzur-Giveon-Kutasov configuration
for $N=1$ duality. The rotation angle  is the $N=2$ 
breaking parameter, the mass of the adjoint scalar in the
$N=2$ vector multiplet. We add some comments on the relevance
of these configurations as  possible stringy proofs of $N=1$ 
duality.        

\Date{March 1996 }


\newsec{Introduction}
Recently, very explicit string realizations of Seiberg's $N=1$ duality
\refs\rsei\ have been proposed in a number of papers. They
involve aspects of D-brane dynamics in non-trivial compactification
manifolds \refs\rvaf, combined with standard $T$-duality, 
 or more complicated  structures in flat
space including both D-branes and $NS$-branes \refs\regk.
A recent work with a unified view is \refs\rvafo.   
 
We study some aspects of the configurations presented
by Elitzur, Giveon and Kutasov (EGK)
 in \refs\regk, which describe a continuous
family of type-IIA brane configurations interpolating between two Seiberg
dual pairs in the simplest case. These manipulations rely heavily on
non-trivial effects of brane dynamics described by Hanany and Witten
(HW) in \refs\rhw. In this note, we exhibit a family of rotated
brane configurations interpolating between a type-IIA four-dimensional
 analogue of the HW 
configurations, and the EGK configuration. This family of configurations
with four-dimensional $N=1$ supersymmetry is a microscopic model
for the simplest deformation of four-dimensional $N=2$ QCD into 
$N=1$ QCD, by giving an $N=1$ preserving 
 mass to the adjoint chiral superfield in the
$N=2$ vector multiplet.  In this way, we make contact with previous
work of Argyres, Plesser and Seiberg in ref. 
\refs\rdeff.

\newsec{Interpolating between the HW and EGK Configurations} 

We will consider the basic set-up of ref. \refs\regk\ in type-IIA
string theory: a configuration containing an $NS_5$  five-brane localized
in the $(x^6, x^7, x^8, x^9 )$ directions, a second $NS_5'$ five-brane
localized in $( x^4, x^5, x^6, x^7) $, at the same value of $x^7$
as the $NS_5$ five-brane, and separated by 
an interval $L_6$ in the $x^6$
direction. We also have a Dirichlet four-brane $D_4$ with world-volume
along $(x^0, x^1, x^2, x^3, x^6)$, stretched in the $x^6$ direction
between the $NS_5$ and $NS_5'$ five-branes. Finally, we have a Dirichlet
six-brane $D_6$  localized in $(x^4, x^5, x^6)$. If we arrange $N_c$
coincident four-branes and $N_f$ six-branes, the previous configuration
defines an $N=1$ super-QCD with gauge group 
 $U(N_c)$  and $N_f$ flavours of quarks
in the fundamental representation, along the four non-compact dimensions
of the $D_4$ world-volume: the space $(x^0, x^1, x^2, x^3)$.    

The amount of supersymmetry is easily characterized. 
In terms of the ten-dimensional chiral and anti-chiral type-IIA spinors:
$\ep = \Gamma^0 \cdots \Gamma^9 \ep$, $\epb = -\Gamma^0 \cdots \Gamma^9 
\epb$, each $NS$-brane imposes the projections   
\eqn\susns{ \ep = \Gamma_{NS}\, \ep \,\,\,\,,\,\,\,\,\,\,\,
 \epb = \Gamma_{NS} \,\epb,}
where $\Gamma_{NS}$ is the product of Dirac matrices along the
brane world-volume directions.  On the other hand, D-branes relate 
both ten-dimensional spinors by the constraint     
\eqn\susd{\epb =\Gamma_D \, \ep.}   
In the  above configuration, the first five-brane $NS_5$ preserves 
$1/2$ of the original ten-dimensional $N=2$ supersymmetry. The second
$NS_5'$ breaks $1/2$ of the remaining supersymmetry, as  
does the $D_6$ brane. These conditions leave four real charges or
$N=1$ in four dimensions. The $NS_5$ and $D_6$ conditions imply the
relation  
\eqn\susif{ \epb= \Gamma^0 \Gamma^1 \Gamma^2 \Gamma^3 \Gamma^6 \, \ep,}
so that another  $D_4$ brane is allowed, extended in
the $(x^0, x^1, x^2, x^3, x^6)$ directions, without any further
breaking of supersymmetry. From the geometry of the configuration
this means that the $D_4$ must stretch between the $NS_5$ and the
$NS_5'$, and be localized at the fixed common $x^7$ position.   

It is easy to see that replacing the $NS_5'$ by a second, displaced 
$NS_5$ leads to $N=2$ supersymmetry on the non-compact part of the
$D_4$ world-volume. This is a result  of eq. \susif\ being
a consequence of \susns\ and \susd\ for the $NS_5$ and $D_6$ branes.  
If we take the five-branes as rigid static objects for the purposes
of defining the effective physics on the $D_4$ world-volume, the
extra scalars in the adjoint representation required by $N=2$
supersymmetry appear because  the $D_4$ is now free to fluctuate in
the $(x^4, x^5)$ plane. So, we have $N=2$ super-QCD with $N_c$
colours and  $N_f$ flavours in a four-dimensional type-IIA 
generalization of
the Hanany-Witten configurations\foot{The four-dimensional configurations
follow from the ones considered in \refs\rhw\ by a $T$-duality in the
$x^4$ direction, under the assumption that the $NS$-branes are inert
under this transformation. Considered as a closed string background,
the string metric component of the type-IIA five-brane has $g_{44} =1$
when the $x^4$ dimension belongs to the world-volume. Therefore, it is
unchanged by
$T$-duality $g_{44} \rightarrow 1/ g_{44}$, and we end
up with a family of type-IIB configurations as in ref. \rhw, averaged
over the compact $x^4$-circle.}.      

This situation immediately suggests   an interpolation between both
types of configurations, by simply rotating the second $NS_5$ into
the $(x^8, x^9)$ plane, to define an $NS_5'$ brane.    
Such a rotation can be performed without breaking all the supersymmetries,
according to the results of ref. \refs\rangles.
The condition is that it can be written as an $SU(n)$ rotation  
for an appropriate  complexification of space.   

 Define the complex
planes $z=x^4 +ix^8$, $w= x^5 +i x^9$. Then, the $NS_5$ is stretched 
in the plane ${\rm Im}\,z = {\rm Im}\,w =0$, whereas the final $NS_5'$ 
configuration lies on ${\rm Re}\,z = {\rm Re}\,w =0$. Clearly, the rotation
\eqn\rot{ z\rightarrow e^{i\theta} z \,\,\,,\,\,\,\,\,w\rightarrow 
e^{-i\theta} w }
is in $SU(2)$ and leaves some unbroken supersymmetry. Since the starting
configuration has four-dimensional $N=2$ supersymmetry, and the final
one at $\theta = \pi/2$ has $N=1$, the minimal amount in four dimensions,
we know that all rotated branes $NS_5^{\theta}$ leave exactly $N=1$
supersymmetry on the $D_4$ world-volume.  
We can see this more explicitly by using \susns--\susif. Defining
\eqn\gammc{a_z = \half (\Gamma^4 +i \Gamma^8) \,\,\,,\,\,\,\,\,
a_w =\half ( \Gamma^5 +i \Gamma^9),}
the condition for unbroken supersymmetry at an angle $\theta$ becomes
\eqn\cond{ (a_z + a_z^{\dagger})(a_w + a_w^{\dagger})\,\ep = 
(e^{i\theta} a_z + e^{-i\theta} a_z^{\dagger}) (e^{-i\theta} a_w +
e^{i\theta} a_w^{\dagger})\,\ep,}  
and both the vacuum $|0\ket$ and the top state $a_z^{\dagger} a_w^{
\dagger} |0\ket$ of the system of two oscillators survive. Moreover,
they have the same ten-dimensional chirality. We can take any of the
two states to build spinors out to the rest of Dirac matrices. We
have six extra Dirac matrices which give a total of $2^{3} $ states.
These are reduced by a factor of $1/4$ by 
 the $NS_5$ and $D_6$ conditions, leaving  two states on top of  each 
of the $z$--$w$ vacua.
 In all, we have four states, corresponding to $N=1 $ in
four dimensions.  

The starting configuration at $\theta=0$ with $N=2$ supersymmetry 
contains an adjoint scalar coming from fluctuations of the $D_4$ in
the $(x^4, x^5)$ plane, $\Phi = X^4 -i X^5 $, where $X^{4,5} $ represent
the $N_c \times N_c$ D-brane position matrices.    On the other hand,
the final EGK configuration at $\theta = \pi/2$ has no scalar moduli,
under the assumption of rigidity of the background branes. Therefore,
it is natural to interpret the rotation angle $\theta$ as a mass parameter
for the $N=2$ adjoint field, inducing a superpotential of the form
$W_{\mu} = \mu\, {\rm Tr} \Phi^2$. For a more precise statement we need
some discussion on the rigidity of the  background branes.   

\newsec{Brane Angles and $N=2$ Breaking Mass} 
A basic assumption of the constructions in \refs\rhw\ and \refs\regk\
is the rigidity of the background branes. In other words, one never  
considers scalar moduli corresponding to $D_4$ fluctuations in the
transverse directions common to both the $D_4$ and the background branes.
We can characterize this rigidity at a quantitative level by adding
convenient mass terms for the corresponding scalar fields. For example,
in the $\theta=0$ configuration, we would ``freeze" the transverse
fluctuations in the $(x^8,x^9)$ plane $\Phi' = X^8 +i X^9$, 
by giving them 
 a large mass $\mu_0 $, at  
 the $D_4$ end-points attached to the $NS_5$ branes. The  full
five-dimensional action on the $D_4$ world-volume  takes the form      
\eqn\m{\eqalign{ S_{5d} =& \int d^4 x\,dx^6 \, \CL_{\rm bulk} + 
\mu_0 \, \int_{x^6 =0} d^4 x \,d^2\theta\, {\rm Tr}\left(\Phi' (x^6 =0)
\right)^2 + \cr 
+ & \mu_0 
\, \int_{x^6 =L_6} d^4 x \,d^2\theta\, {\rm Tr} \left( \Phi' (x^6 = L_6)
\right)^2 \,\,\,+ {\rm h.c.} } }  
After dimensional reduction at small $L_6$ we just keep zero modes in
the $x^6$ direction and then $\Phi' (x^6 =0) = \Phi' (x^6 =L_6)$. We end
up with 
\eqn\four{ S_{4d} = L_6 \int d^4 x\, \CL_{\rm bulk} + 2\mu_0 \,  
 \int d^4 x \,d^2\theta\, {\rm Tr}(\Phi')^2 \,\,\,+{\rm h.c.} }  
In the decoupling limit $\mu_0 \rightarrow\infty$, the $\Phi'$ fields
are frozen\foot{A dynamical motivation for the rigidity of the $NS$ branes
as  compared to the $D_4$ branes could be found in the parametrically
larger tension, at weak coupling $T_{NS} \sim g_{\rm st}^{-2} \gg 
g_{\rm st}^{-1} \sim T_D$. },    
  and we are left with the $N=2$ four-dimensional theory, with
bare gauge coupling $g_{\rm bare} \sim L_6^{-1/2}$.

It is now very easy to incorporate the rotation of the second $NS_5$.
We simply modify the boundary action at $x^6 = L_6$, by writing
a superpotential
\eqn\rotb{ W_{\theta} (x^6 = L_6) = \mu_0 \,
{\rm Tr} (\Phi'_{
\theta})^2 ,}
with $\Phi'_{\theta} = X_{\theta}^8 + i X_{\theta}^9$, and 
\eqn\defx{\eqalign{X_{\theta}^8 =& X^4 \,{\rm sin}\theta  +X^8\,
 {\rm cos}\theta
 \cr
 X_{\theta}^9 =& X^9\,{\rm cos}\theta  - X^5\, {\rm sin}\theta  . } } 
Working out the dimensional reduction we find the following superpotential
in four dimensions:
\eqn\fursup{ W_{\theta} = \mu_0 \, (1+{\rm cos}^2
 \theta)\, {\rm Tr} (\Phi'  
)^2 + \mu_0 \, {\rm sin}^2 \theta \, {\rm Tr} (\Phi)^2 
+ \mu_0 \, {\rm sin} 2\theta \, {\rm Tr}\, (\Phi\,\Phi'). } 
Thus, after diagonalization for small $\theta$, there is a heavy field
with mass of order $\mu_0$, and a light field with mass parameter
\eqn\mpa{\mu = \mu_0 {{\rm sin}^2 \theta \over 1 + {\rm cos}^2 \theta} 
\sim {\mu_0 \over 2}\theta^2 . }
For $\theta \sim \pi/2$ the two fields are  decoupled, as corresponds
to the absence of moduli in the EGK configuration.     
    
The ``duality trajectory" of brane configurations described in 
\refs\regk\ is easily generalized to the rotated configurations, as the
corresponding intermediate Higgs phases with a non-zero Fayet-Iliopoulos
coupling (FI) do exist for the deformed $N=2$ theories. 
Indeed, for $\theta=0$, the EGK trajectory realizes explicitly an  
``$N=2$ duality" between $U(N_c)$ and $U(N_f -N_c)$ theories, similar
to the one described in \rvaf, \rvafo, \rdosdual.    

 The final configuration
obtained after passing through the Higgs branch with a non-zero FI
term consists of $N_f -N_c$ $D_4$ branes stretched in the $x^6$ between
two parallel $NS_5$ branes, together with $N_f$ $D_4$ branes stretched
between the second $NS_5$ and $N_f$ $D_6$ branes\foot{This is not
an $s$-configuration, in the terminology of ref. \rhw, because the
local linking of $D_4$ branes to $D_6$ branes is one to one.}. The
fundamenal type-IIA strings stretching between $D_4$ branes on both
sides of the second $NS_5$ provide the $N_f$ massless quark flavours.
Notice that there are no extra ``magnetic mesons" in this $N=2$
 configuration, since the $N_f$ $D_4$ branes between the second $NS_5$
and the $D_6$ branes  are rigid. This is, however, a subtle point, since
one might argue that a flavour gauge group $U(N_f)$ should be present,
with inverse squared coupling proportional to the $x^6$-distance between 
the $D_6$ branes and the $NS_5$ brane. Then, $N=2$ supersymmetry 
 would imply the existence of  an adjoint
superfield for the  flavour group with $N=2$ couplings, which would 
naturally qualify for Seiberg's magnetic mesons. 
  This is a subtle question because
of the very particular structure of $D_4$--$D_6$ linking (one to one).  
In any case, if we stick to the convention that background branes
are rigid for the purposes of defining massless dynamics on the
$D_4$ world-volume,  there is apparently no room for flavour
gauge group in the $N=2$ version of the final EGK configuration. 

In ref. \rvafo, a similar arrangement of branes was proposed, realizing
an $N=1$ duality trajectory. The starting configuration is an
 $NS_5'$ brane
connected to an $NS_5$ brane by $N_c$ $D_4$ branes stretched in the
$x^6$ direction, which in turn is further connected by $N_f$ $D_4$ branes
to a second $NS_5'$ brane. The duality trajectory proceeds by switching
the positions of the first   $NS_5'$ and the middle $NS_5$;  
 the change of gauge group comes about by
 reconnection of branes at the middle background brane.  
In this construction, there is an obvious flavour--colour symmetry from
the geometry of the configurations, and we are clearly describing
$U(N_c)\times U(N_f)$ or $U(N_f -N_c)\times U(N_f)$ gauge theory 
 with matter
in the $(N_c, N_f) + {\rm h.c.}$. In this context,
 regarding $U(N_f)$ as a global
 flavour group
is more a question of making its gluous very weakly coupled by adjusting
the brane distances.      The rotated configurations
considered in this paper can be trivally extended to this case: by
a complex rotation of the middle $NS_5$ into the $(x^8, x^9)$ directions
we achieve an $N=2$ configuration with three parallel $NS_5'$ branes
connected by $D_4$ branes as above. Here we $do$ find the appropriate 
adjoint scalars required by $N=2$ supersymmetry 
 both in the ``colour" and ``flavour" sectors, because both sets of
$D_4$ branes are free to fluctuate in the $(x^8,x^9)$ directions.

Coming back to the EGK configurations, the 
 effect of turning on a rotation angle of the middle $NS_5$ is again
a soft $N=1$ mass $\mu \sim \theta^2$ for the adjoint $\Phi=X^4 -iX^5$.
At the same time, the inverse effect occurs as $\theta\rightarrow \pi/2$,
near the EGK configuration. Now the same analysis applied to the
$N_f$ ``flavour" $D_4$ branes leads to a mass $\mu_f \sim (\theta - \pi/
2)^2$ for the ``flavour adjoint" $\Phi_f' = X_f^8 +iX_f^9$, where $X_f^{8,9}$
denote the corresponding position matrices of the $N_f$ $D_4$ branes
stretching from $NS'_5$ and the $N_f$ $D_6$ branes.
 These are just
Seiberg's extra magnetic mesons coming down as $
\theta\rightarrow \pi/2$.  

Thus, the complete picture in the $(g_{\rm bare} \sim  L_6^{-1/2} , \mu)$
plane is strongly reminiscent of similar discussions in the field
theoretical context of ref. \refs\rdeff.    

\newsec{Concluding remarks}
In this note we have shown that $N=2$ and $N=1$ brane configurations,  
appropriate for discussions of four-dimensional duality, can be connected
by a rotation process of one of the branes. This realizes the simplest
deformation on $N=2$ QCD by the lifting of the adjoint chiral superfield. 
It would be very interesting to sharpen the analogy between this 
two-parameter family of theories and the analogous treatment in
ref. \refs\rdeff. These authors pinned the degrees of freedom of the
magnetic dual by slightly breaking $N=2$ to $N=1$ with $\mu \ll        
\Lambda_{N=2}$, the key point being that vacua at the roots of the
Higgs branches with the right properties are not lifted. Then, deforming
the theory by increasing $\mu$ past $\Lambda_{N=2}$, one finds the
microscopic electric description. This process is clearly analogous
to the interplay between brane motion (variation of the bare couplings),  
and angle rotation ($N=1$ breaking mass).  

There are some superficial differences, though. For example, in the
field theoretical treatment of \refs\rdeff, the dual quarks and 
gluons and  the magnetic mesons 
 evolve from vacua at the baryonic and
non-baryonic roots respectively. The distance between these roots is
of order $\Lambda_{N=2}$, the strong interaction scale, 
 which vanishes in the decoupling limit
$\mu\rightarrow\infty$ with $\Lambda_{N=1}^{3N_c - N_f} = \mu^{N_c} 
\Lambda_{N=2}^{2N_c - N_f} $ fixed, thereby merging into a single
vacuum. In the brane treatment, however, the magnetic mesons 
already appear at a microscopic level. In this sense, it is interesting
to note that they are absent for the $N=2$ configuration, perhaps
in analogy with the previously mentioned 
 low-energy splitting between baryonic
and non-baryonic roots.    

A more explicit connection with ref. \refs\rdeff\ might be 
achieved by regarding the bare coupling of the effective four-dimensional
theory fixed at the string scale $g_{\rm bare}^2 \sim g_{\rm st}$, and
considering the physics at a scale $M$, with $M\,L_6 \ll 1 $ kept
fixed as we move $L_6$. Then, brane motion really corresponds to
renormalization group flow, by changing the scale $M$. Taking $M$
to the infrared, and at the same time deforming the theory to
$\mu\rightarrow \infty$ at different relative velocities, identifies
both Seiberg duals at intermediate scales. The full field theoretical
analysis is recovered in the complete decoupling limit for infinitely
rigid branes; according to \mpa, we take $\mu_0 \rightarrow\infty$
and $\theta\rightarrow 0$, keeping $\mu$ fixed. However, in this
way we lose the $\theta\sim \pi/2$ region and the stringy characterization
of the magnetic mesons. So, it might be useful to keep a finite
$\mu_0$ after all.

  The requirement of having to pass
 through the Higgs branch,  
in order to avoid an infinite coupling singularity in the stringy     
setting, could be related to the fact that, as we reduce $L_6$  
and take $M$ past $\Lambda_{N=2}$ into the infrared, the baryonic 
Higgs cone splits from the classical unbroken $SU(N_c)$ vacuum. Therefore,
we need to go through the Higgs branch in order to reach the 
infrared-free $SU(N_f - N_c)$ vacuum, unless we take the ``short-cut"
through the Coulomb phase, which corresponds in the brane language
to the relative  splitting of the $N_c$ 
$D_4$ branes in the $(x^4, x^5)$ directions. In this respect, a
 disturbing feature
of the brane configurations is the fact that one always gets unitary
gauge groups, rather than special unitary gauge groups. As pointed
out in \refs\rhw, this means that baryon number is effectively gauged, and
the baryonic branch is not present in the $U(N_c)$ moduli spaces.  
In order to fully compare the brane approach to the discussion in
\refs\rdeff, one should somehow restore the baryonic Higgs branch
in the brane picture.

These analogies should become more specific. This is an
important point in elucidating the $N=1$ duality mapping, because
the continuous family of brane configurations interpolating between
dual pairs does not guarantee the infrared equivalence of the two  
theories \refs\rvafo.
 Indeed, they are clearly  inequivalent along the $\theta=0$,  
$N=2$ slice, which connects an asymptotically free theory with an
infrared-free theory for $N_c \le N_f \le 2N_c$.
 Therefore, it is unlikely that purely microscopic
considerations will qualify for an unambiguous proof of $N=1$ duality, 
and some low-energy input, in the spirit of \refs\rdeff, will be
necessary.       
      
\newsec{Acknowledgements}
It is a pleasure to thank L. Alvarez-Gaum\'e and A. Schwimmer for
useful discussions.   

\listrefs
\bye